\documentclass[iop]{emulateapj}
\pdfoutput=1 

\begin{document}

\title{BLAZAR RADIO AND OPTICAL SURVEY (BROS): A CATALOG OF
  BLAZAR CANDIDATES SHOWING FLAT RADIO SPECTRUM AND THEIR
  OPTICAL IDENTIFICATION IN PAN-STARRS1 SURVEYS}

\author{Ryosuke Itoh} 
\affiliation{Bisei Astronomical Observatory,  1723-70 Ohkura, Bisei-cho, Ibara, Okayama, 714-1411, Japan}

\author{Yousuke Utsumi}
\affiliation{Kavli Institute for Particle Astrophysics and Cosmology (KIPAC), SLAC National Accelerator Laboratory, Stanford University, 2575 Sand Hill Road, Menlo Park, CA 94025, USA}

\author{Yoshiyuki Inoue}
\affiliation{Interdisciplinary Theoretical \& Mathematical Science Program (iTHEMS), RIKEN, 2-1 Hirosawa, Saitama 351-0198, Japan}
\affiliation{Kavli Institute for the Physics and Mathematics of the Universe (WPI), UTIAS, The University of Tokyo, Kashiwa, Chiba 277-8583, Japan}

\author{Kouji Ohta}
\affiliation{Department of Astronomy, Kyoto University,  Kyoto 606-8502,
Japan}

\author{Akihiro Doi}
\affiliation{The Institute of Space and Astronautical Science, Japan Aerospace Exploration Agency, 3-1-1 Yoshinodai, Chuou-ku, Sagamihara, Kanagawa 252-5210}

\author{Tomoki Morokuma}
\affiliation{Institute of Astronomy, Graduate School of Science, The University of Tokyo, 2-21-1 Osawa, Mitaka, Tokyo 181-0015, Japan}

\author{Koji S. Kawabata}
\affiliation{Hiroshima Astrophysical Science Center, Hiroshima University, 1-3-1 Kagamiyama, Higashi-Hiroshima, Hiroshima 739-8526, Japan}
\affiliation{Department of Physical Science, Hiroshima University, 1-3-1 Kagamiyama, Higashi-Hiroshima, Hiroshima 739-8526, Japan}
\affiliation{Core Research for Energetic Universe (Core-U), Hiroshima University, 1-3-1 Kagamiyama, Higashi-Hiroshima, Hiroshima 739-8526, Japan}

\author{Yasuyuki T. Tanaka} 
\affiliation{Hiroshima Astrophysical Science Center, Hiroshima University, 1-3-1 Kagamiyama, Higashi-Hiroshima, Hiroshima 739-8526, Japan}

\begin{abstract}
Utilizing the latest and the most sensitive radio and optical catalogs, we completed a new blazar candidate catalog, Blazar Radio and Optical Survey (BROS), which includes 88,211 sources located at declination $\delta > -40^{\circ}$ and outside the galactic plane ($|b| > 10^{\circ}$). We list compact flat-spectrum radio sources of $\alpha > -0.6$ ($\alpha$ is defined as $F_{\nu} \propto \nu^\alpha$ ) from 0.15~GHz TGSS and 1.4~GHz NVSS catalogs. We further identify optical counterparts of the selected sources by cross-matching with Pan-STARRS1 photometric data. Color-color and color-magnitude plots for the selected BROS sources clearly show two distinct populations, An "quasar-like" population consisting of both flat-spectrum radio quasars and BL Lac type objects. On the other hand, an "elliptical-like" population of mostly BL Lac-type objects is buried in the elliptical galaxy.
The latter population is also reported in previous catalogs, but BROS catalog brought new larger sample of these population, due to the lower radio flux threshold of our selection. Model calculations show that the "elliptical-like" population consists of elliptical galaxies located at redshift z $\lesssim$ 0.5, which is also supported by the logN-logS distribution of the power-law index of $1.49 \pm 0.05$. This BROS catalog is useful for identifying the electromagnetic counterparts of ultra-high-energy cosmic rays and PeV neutrinos recently detected by IceCube, as well as nearby BL Lac objects detectable by future high sensitivity TeV telescopes, such as Cherenkov Telescope Array.

\end{abstract}


\section{Introduction}
The extragalactic gamma-ray sky is dominated by blazars \citep[e.g.,][]{TeVCat,2015ApJS..218...23A}. Blazars are one population of active galactic nuclei (AGN) whose jets are directed toward the Earth. There are 3,131 blazars in the latest {\it Fermi} 8-year catalog \citep[][4FGL catalog]{2019arXiv190210045T},  which constitutes 61.8\% of the total number of detected samples, 5,065 sources. In the 4FGL catalog, however, 657 sources (13.0\%) at high galactic latitude $|b| > 10^{\circ}$ are still unassociated, despite their location. The nature of these sources has yet to be resolved; however, their locations and multi-wavelength properties suggested that most of them are blazars \citep[e. g.,][]{2019MNRAS.486.3415L,2020arXiv200106010Z}. To firmly identify them as blazars, flat spectrum features in the radio band would be required, as one of the most important characteristics of blazars \citep{1995PASP..107..803U}. However, currently available blazar catalogs do not provide such evidence for these high latitude, unassociated objects. Therefore, a new blazar catalog containing radio spectrum information and covering fainter objects is required to elucidate the nature of such {\it Fermi} unassociated objects.

A new deep and wide blazar catalog will be useful for future extragalactic surveys in the TeV band. The next generation imaging atmospheric Cherenkov telescopes (IACTs), namely the Cherenkov Telescope Array (CTA), is expected to be operational in 2022\footnote{\url{https://www.cta-observatory.org}}. CTA will have about 10 times the sensitivity of current IACTs \citep{2019scta.book.....C}. More than 1000 TeV sources are expected to be detected over the entire sky including non-{\it Fermi}-detected blazars such as extreme blazars  \citep[e.g.,][]{Inoue2010,Dubus2013,CTA2019,2019arXiv191105027C,2019arXiv191106680M,2019arXiv190510771T}. The blazar catalog will provide insight into the properties of new TeV blazars.

Such a catalog also helps in the identification of the counterparts of high-energy neutrino events. On 22 September 2017, the IceCube collaboration reported the first candidate of high-energy neutrino event associated with the gamma-ray flare activity of blazar TXS ~0506+056 \citep[IceCube-170922A,][]{2017ATel10791....1T,2018Sci...361.1378I}. For the counterpart identification, optical and multi-frequency follow-up observations have played a critical role. The enhanced activity of TXS~0506+056 was first identified by optical observations targeting blazars in the error circle of IceCube-170922A, which led to the discovery of the associated gamma-ray activity \citep[e.g.,][]{2017GCN.21930....1K,2017GCN.21941....1E,2017ATel10844....1Y,Morokuma2020}. Although covering the entire error circle regions of neutrino events would be the most effective and unbiased, such observations are extremely expensive for standard optical/infrared cameras. In the case of IceCube-170922A, the error region was about $1.6 \times 0.8$ deg$^2$ with a 90\% confidence level, which requires $\sim 50$ pointings and a camera with a field-of-view (FoV) of $10 \times 10$ arcmin$^2$ to cover the entire region. 
An alternative strategy is to observe all blazar candidates within the IceCube error circle, based on the IceCube-170922A event. Although this strategy assumes that neutrinos are emitted by blazars, it allows us to search blazars efficiently, due to the low surface density of blazars $\lesssim1\ {\rm blazar}/{\rm deg}^2$ \citep[e.g.,][]{Inoue2009,Ajello2015,2019A&A...632A..77C}.
In this strategy, pre-identification of blazar candidates across the entire sky is important for efficient follow-up observation. 

Radio sky maps must be used to construct a blazar catalog, as blazars are characterized by their flat radio spectrum \citep{UrryPadovani1995}. In 2017, radio survey data at 0.15~GHz taken with the Giant Metrewave Radio Telescope (GMRT), specifically, the Tata Institute of Fundamental Research GMRT Sky Survey (TGSS), was released \citep{2017AaA...598A..78I}. The TGSS catalog includes 0.62 million sources in the sky above $\delta > -53^{\circ}$ ($3.6\pi$ steradians). By cross-matching with the NRAO Very Large Array Sky survey (NVSS) catalog at 1.4 GHz \citep{1998AJ....115.1693C}, which also covers the sky of $\delta > -40^{\circ}$ and overlaps nearly the same region as TGSS, we can create the largest catalog of flat-spectrum radio sources to date. This effort is expected to compliment previous flat-spectrum radio source catalogs, such as CLASS \citep{2003MNRAS.341....1M,2003MNRAS.341...13B} and CRATES \citep{2007ApJS..171...61H}. However, both the selection frequency and the flux threshold are different from those of previous catalogs, which provides a large sample of flat-spectrum radio sources.

In addition to radio data, optical information will allow us to categorize the properties of blazars, according to their types and host galaxies. Here, the Pan-STARRS1 (PS1) optical survey catalog was released in 2016 December \citep{2016arXiv161205560C}. The PS1 catalog provides the deepest photometric data over the large region of $\delta > -30^{\circ}$ with point source $5 \sigma$ limiting magnitudes of 23.3, 23.2, 23.1, 22.3 and  21.4 mag in {\it g, r,i, z} and {\it y} bands, respectively. By combining the PS1 catalog with the radio catalogs, we performed the most sensitive optical counterpart search for selected flat-spectrum radio sources, i.e., blazar candidates. 

In this paper, we refer to our blazar candidate catalog as Blazar Radio and Optical Survey (BROS). The remainder of this paper is organized as follows. We describe details of our BROS catalog in \S 2. A discussion and conclusions are presented in \S 3 and \S 4, respectively. 


\section{The BROS Catalog}
\subsection{BROS sources selection}
\label{catalog}

\begin{figure*}[!htb]
  \centering
  \includegraphics[angle=0,width=15cm]{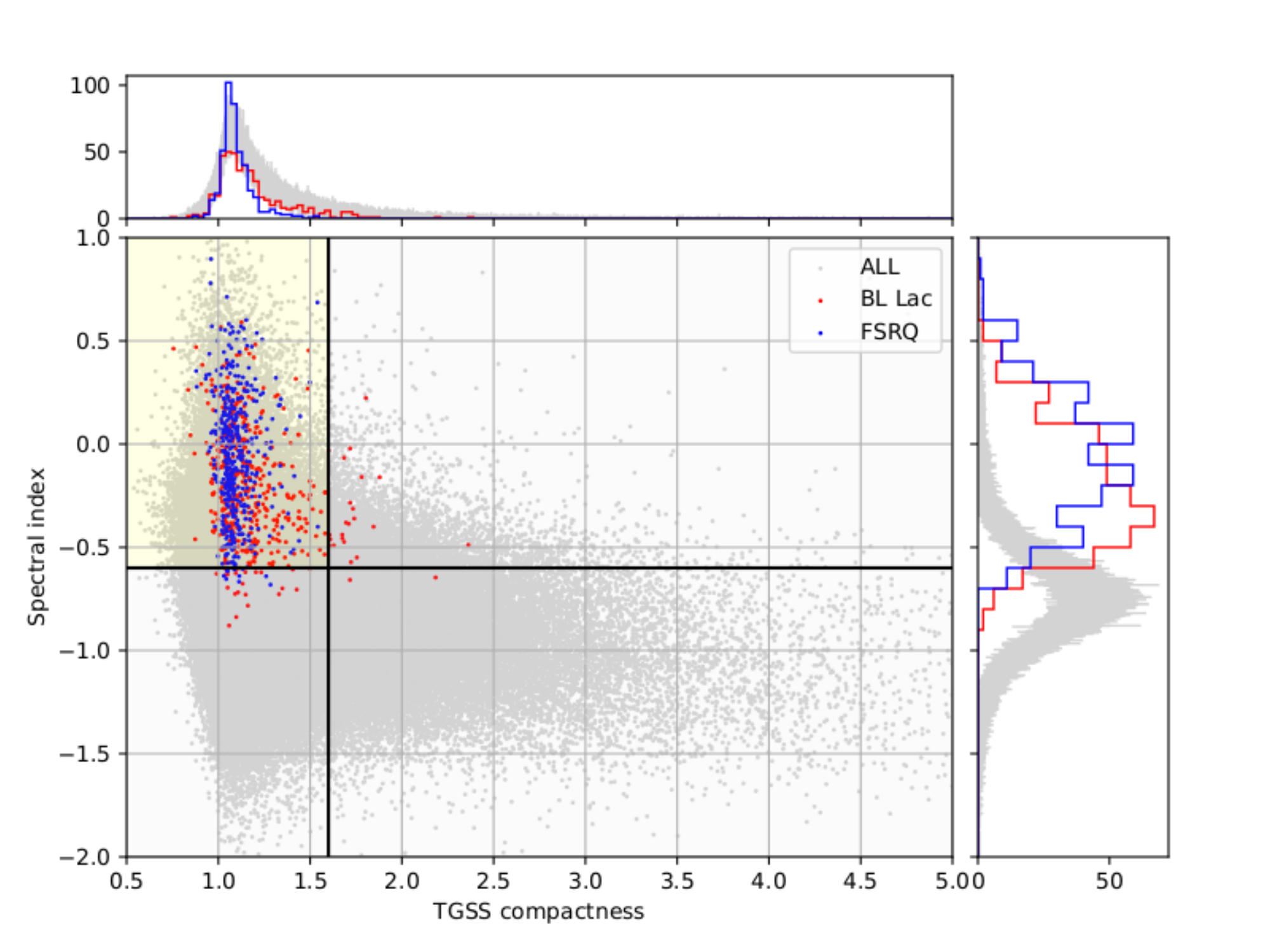}
  \caption{Scatter plot of radio compactness and spectral index in the radio band.
  Red data indicates gamma-ray emitting BL Lac and blue data indicates gamma-ray flat-spectrum radio quasar (FSRQ) detected by Fermi/LAT. Gray data point indicates that all radio sources in TGSS-NVSS catalog. 
  The BROS sources reside in the yellow shaded region.
  }
  \label{fig:SandCplot}
\end{figure*}

To select flat-spectrum radio sources, we use the TGSS-NVSS spectral index maps and catalog \citep[TGSS-NVSS spindx/catalog, ][]{2018MNRAS.474.5008D}, which contains about 0.5 million sources. Fluxes and the spectral index are taken from 0.15~GHz TGSS and 1.4 GHz NVSS catalogs over the wide region of $\delta > -40^{\circ}$. In the TGSS-NVSS catalog, the images of the NVSS and TGSS are re-processed, combined, and then the sources are extracted. Therefore, the TGSS-NVSS catalog differs slightly from the original TGSS and NVSS catalogs \citep[see ][for details]{2018MNRAS.474.5008D}. 
It is estimated that the typical background root mean square (RMS) noise of the NVSS-TGSS catalog is $\sim 0.45$ mJy beam$^{-1}$ at 1.4 GHz and $\sim 8$ mJy beam$^{-1}$ at 0.15~GHz. The RMS noise value at 0.15~GHz is slightly higher than the value in the original TGSS catalog, as TGSS images are convolved to adjust the resolution for NVSS images in the TGSS-NVSS catalog. Mean estimated astrometric error is 3.5 arcsec for the TGSS-NVSS catalog.

From the TGSS-NVSS catalog, we excluded sources located within the galactic plane of $|b| < 10^{\circ}$ where galactic objects are dominant. Then, we selected sources with the source classification flag of 'S' in the TGSS-NVSS catalog, corresponding to the sources detected both in TGSS and NVSS with no other sources in the search region  \citep[see Section 2, ][for details]{2018MNRAS.474.5008D}. 

Among the rest, we selected flat-spectrum and compact objects, i.e., blazar candidates. To define the spectral flatness and the compactness of blazars in this research, we examined known blazars detected by {\it Fermi}/LAT. Figure \ref{fig:SandCplot} shows a scatter plot of the compactness and spectral index of all TGSS-NVSS sources and gamma-ray blazars \citep[4LAC;][]{2019arXiv190510771T}, as shown in the TGSS-NVSS catalog. The spectral index $\alpha$ is defined as $F_{\nu} \propto \nu^{\alpha}$ between the 0.15 and 1.4~GHz. The compactness $C$ is defined as $C \equiv F_{{\rm integral}}/F_{{\rm peak}}$, where $F_{{\rm integral}}$ and $F_{{\rm peak}}$ are the integrated and peak flux measured using TGSS, respectively. We introduce $C$ to avoid contamination of radio galaxies (misaligned blazars). 
As shown in the figure, 94\% of {\it Fermi}-LAT gamma-ray blazars are located in $C\le1.6$ and $\alpha\ge-0.6$. Thus, we selected sources with those criteria for this research.

After applying the selection above, 88,211 BROS sources were left. From Figure \ref{fig:SandCplot}, 58 {\it Fermi}-LAT gamma-ray blazars which include several famous blazars such as 1ES 1215+303 ($z=0.131, C=2.0$) and OQ 530 ($z=0.151, C=1.9$) were excluded.
Twenty-four of these blazars have $C > 1.6$ and mostly reside nearby ($z<0.5$), compared to samples with $C<1.6$.
These sources are actually resolved in TGSS images.
We also note that a small fraction $\lesssim0.5$\% of gamma-ray blazars have steep spectrum lobes at low frequency \citep{2016A&A...588A.141G}. These populations are missed in the BROS catalog.

\subsection{Optical Counterparts}
\label{selection}

\begin{figure}[!htb]
  \centering
  \includegraphics[angle=0,width=8cm]{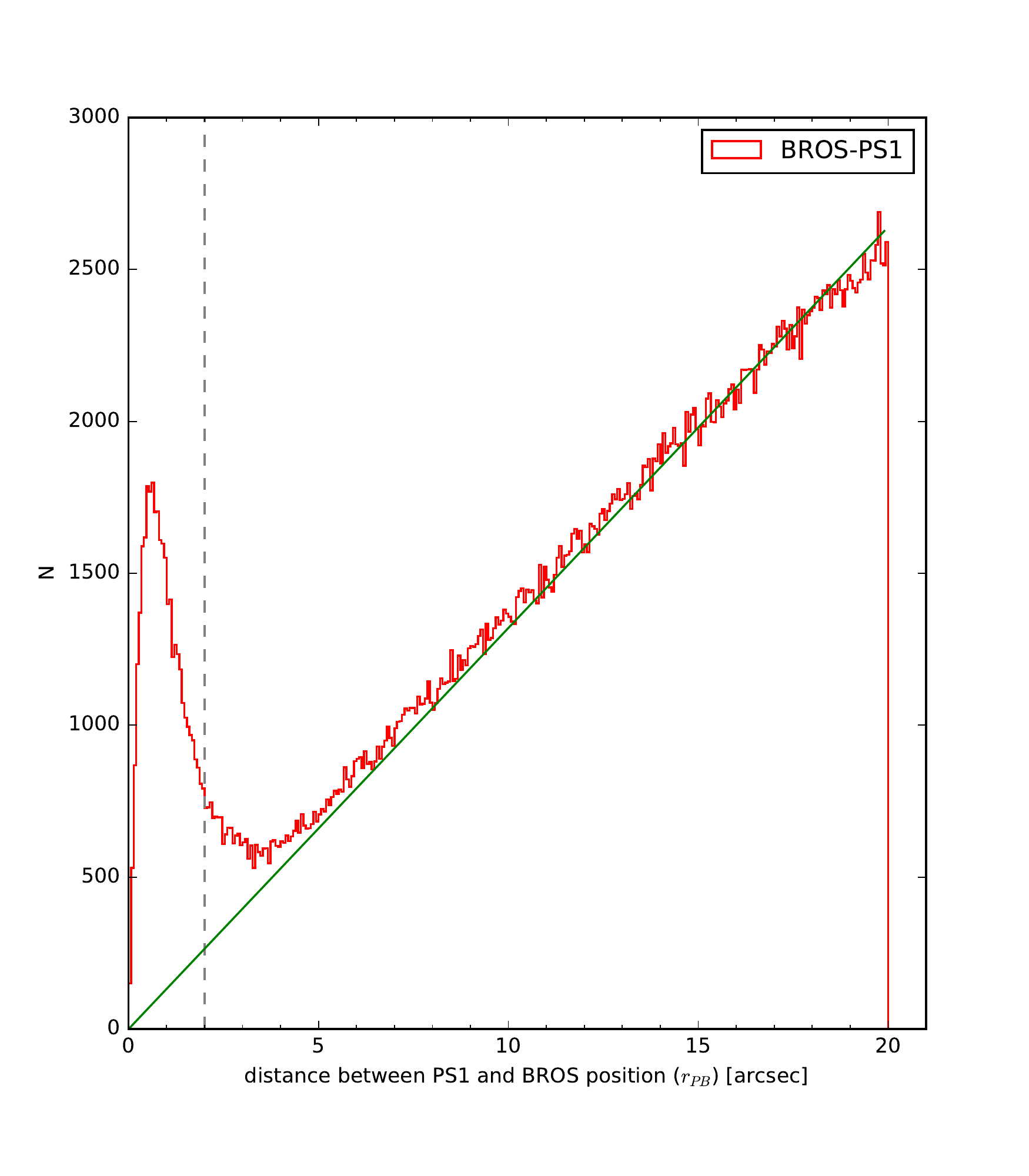}
  \caption{Distance distribution between BROS and PS1 sources. 
  Green line indicates the expected number of association by 
  chance coincidence and dashed line indicates $r_{\rm tol} = 2$ 
  arcsec position (see text for detail). 
  }
  \label{fig:distance_BROS_PS1}
\end{figure}

To search for the optical counterparts of the selected flat-spectrum radio sources, we utilized the Pan-STARRS photometric catalog\footnote{available at \url{http://mastweb.stsci.edu/ps1casjobs/home.aspx}} \citep[][]{2016arXiv161205560C}. The covered region of the PS1 data ($3\pi$ sky above $\delta > -30^{\circ}$ ) matches that of the TGSS-NVSS catalog. The PS1 data were taken between 2010 and 2012 with a 1.8~m telescope near the summit of Haleakala in Hawaii.

Using the PS1 photometric catalog, we identified the optical counterparts of BROS sources by finding the closest PS1 neighbor within a tolerance radius $r_{\rm tol}$. First, we collected all PS1 sources detected in $r$ band within 20 arcsec of the selected flat-spectrum radio sources, which is large enough compared to positional uncertainty in the radio band (typically a few arcseconds).Figure \ref{fig:distance_BROS_PS1} shows the number count of all PS1 neighbors to BROS sources in an annulus between $r_{PB}$ and $r_{PB} + dr_{PB}$ (we set $dr_{PB} = 0.066$ arcsec to reduce uncertainties of each bin less than 5\%).
The number count peaks at 0.5 arcsec and increases again toward larger radii. The first peak is mostly dominated by a true-matched source, while the latter increse is caused by unrelated sources randomly coincident, $\propto \pi n r_{PB}dr_{PB}$, where $n$ is the number density of PS1 sources and $r_{PB}$ is the separation of neighbors with respect to a BROS source. The green line in Figure \ref{fig:distance_BROS_PS1} indicates the linear line representing the association by chance between random positions and PS1 sources. 
This line is a measure of "the background". The tolerance radius $r_{\rm tol}$ is determined empirically to minimize contamination by chance as well as to maximize the true-matched sources. We adopted $r_{\rm tol}$ = 2 arcsec.
Finally, we obtained 42,757 BROS optical counterparts. From Figure \ref{fig:distance_BROS_PS1}, we estimated the contamination ratio $\sim$ 10\% and overlooked ratio $\sim$ 19\% for our data selection.

\begin{figure}[t]
  \centering
  \includegraphics[angle=0,width=8cm]{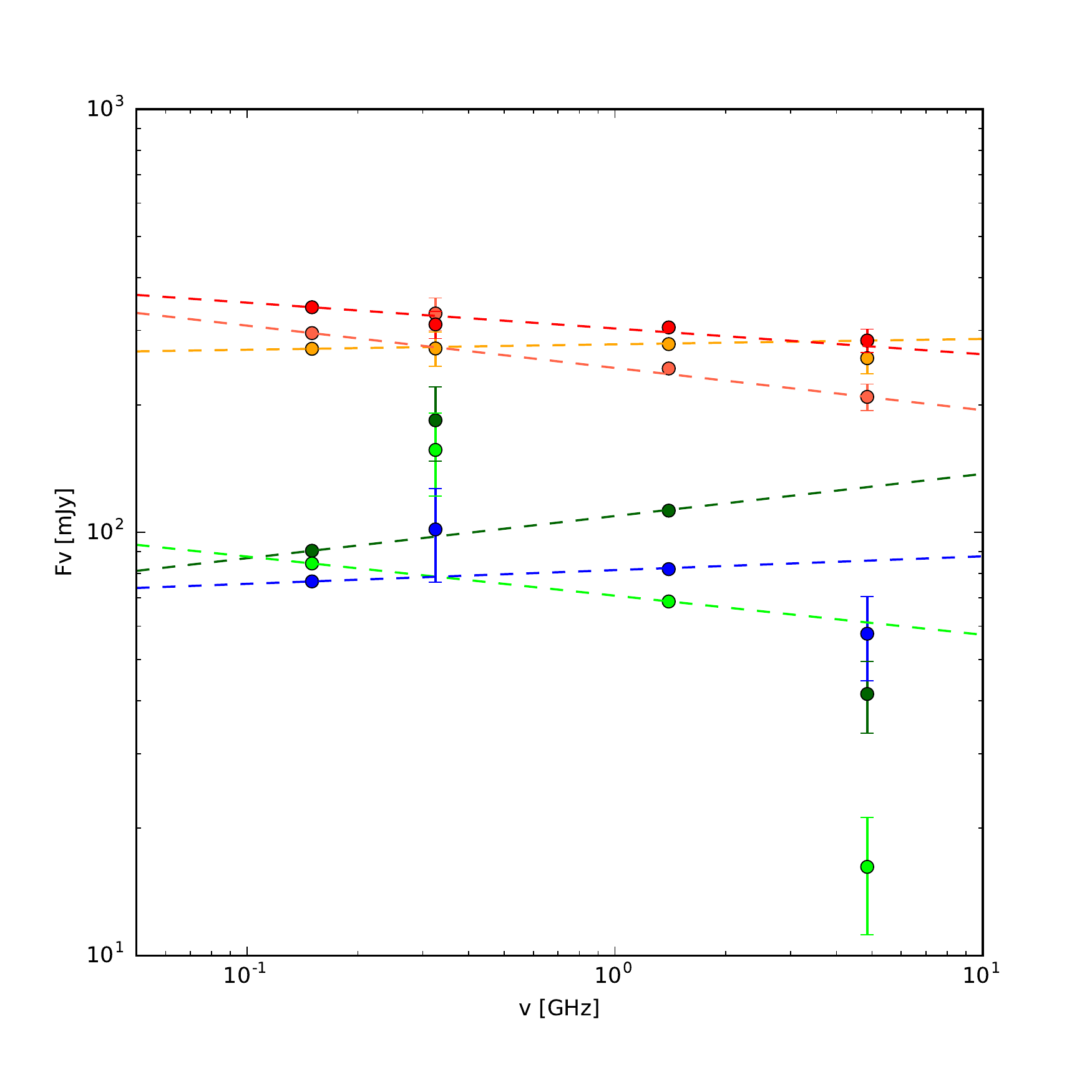}
  \caption{Sample spectra for GPS and flat spectrum sources. Upper three sources with reddish data points indicates the sources showing flat spectrum. Lower bluish data points correspond to GPS objects having bent spectra. Dashed line indicates the estimated spectra derived by TGSS-NVSS spectral index at 0.15~GHz and 1.4~GHz.}
  \label{fig:GPS_example}
\end{figure}

\subsection{Contamination of the gigahertz peaked-spectrum and compact steep-spectrum radio sources}

Gigahertz peaked-spectrum (GPS) and compact steep-spectrum (CSS) radio sources are a kind of compact AGN. Given that we selected sources using only two bands (0.15 and 1.4~GHz), the sources with a spectral peak at $\sim$~GHz can be misidentified as flat spectrum objects in the BROS. 

To investigate the contamination fraction of GPS and CSS radio sources, we checked the multi-band spectrum in the radio band, taking the cross-match between BROS and the multi-frequency radio source catalog by \citet[][hereafter, Kimball catalog]{2014arXiv1401.1535K}, which contains flux values from GB6 at 4.85~GHz and WENSS at 325~MHz \citep{1997A&AS..124..259R,1996ApJS..103..427G}. This enables the blazars and GPS/CSS sources to be distinguished by checking the existence of a peak at $\sim$~GHz. 

Figure \ref{fig:GPS_example} shows examples of GPSs and blazars. We estimated the expected GB6 and WENSS flux densities using only NVSS flux and TGSS-NVSS spectral index (i.e., assuming a single power-law spectrum) and comparing it with the flux reported in the Kimball catalog.  $\sim 10,000$ BROS sources have a counterpart in the GB6 and/or WENSS catalogs. About 25\% of them are identified as GPSs whose source flux is deviates from the flat spectrum at the $\ge$5 $\sigma$ confidence level. For those objects, we added a flag of  "G" to the "SourceFlag" column. However, the GB6 and WENSS catalogs do not cover all-sky condition and have different sensitivity compared to the TGSS-NVSS catalog. Therefore most of the BROS samples ($\sim 70,000$ sources) do not have counterparts in the Kimball catalog. These sources are treated as "blazars" in the BROS catalog; however, the GPS/CSS sources may be contaminated.

\begin{deluxetable*}{lll}[t]
 \tablecaption{The BROS catalog description \label{tab:bros_ps1}}
 \tablecolumns{3}
 \tablehead{
 \colhead{Column name} & \colhead{Format} & \colhead{Notes} 
 }
\startdata
BROS-ID    & -      & source ID\\
spindx-RA  & degree & RA from TGSS-NVSS catalog$^{a}$ \\
spindx-Dec & degree & Decl. from TGSS-NVSS catalog$^{a}$ \\
spindx-TGSS-Flux       & Jy     & TGSS Total flux$^{a}$ \\
spindx-TGSS-Flux-error & Jy     & TGSS Total flux error$^{a}$ \\    
spindx-NVSS-Flux       & Jy     & NVSS Total flux$^{a}$ \\
spindx-NVSS-Flux-error & Jy     & NVSS Total flux error$^{a}$ \\
spindx-Spectral-Index  & -      & Spectral index between 0.15~GHz and 1.4 GHz$^{a}$ \\
spindx-Spectral-Index-error  & - & Spectral index error$^{a}$ \\
SourceFlag & strings & Source type flag, "Q", "E", "G" or "B" \\
radio-Compactness & -- & Radio compactness defined as total flux / peak flux (NVSS) \\
radio-Compactness-error & -- & Radio compactness error \\
optical-Compactness & -- & Optical compactness defined as $M_{\rm{psf}} - M_{\rm{Kron}}$ (PS1, i-
band) \\
optical-Compactness-error & -- & Optical compactness error \\
PS1-BROS-separation & arcsec & source separation angle between TGSS-NVSS and PS1 catalog 
position\\
PS1-r-RA   & degree    & {\it r}-band R.A. position$^{b}$ \\
PS1-r-Dec  & degree    & {\it r}-band Decl. position$^{b}$ \\
PS1-g-Kronmag       & magnitude & {\it g}-band Kron magnitude$^{b}$\\
PS1-g-Kronmag-error & magnitude & {\it g}-band Kron magnitude error$^{b}$\\
PS1-g-PSFmag        & magnitude & {\it g}-band PSF magnitude$^{b}$\\
PS1-g-PSFmag-error  & magnitude & {\it g}-band PSF magnitude error$^{b}$\\
PS1-r-Kronmag       & magnitude & {\it r}-band Kron magnitude$^{b}$\\
PS1-r-Kronmag-error & magnitude & {\it r}-band Kron magnitude error$^{b}$\\
PS1-r-PSFmag        & magnitude & {\it r}-band PSF magnitude$^{b}$\\
PS1-r-PSFmag-error  & magnitude & {\it r}-band PSF magnitude error$^{b}$\\
PS1-i-Kronmag       & magnitude & {\it i}-band Kron magnitude$^{b}$\\
PS1-i-Kronmag-error & magnitude & {\it i}-band Kron magnitude error$^{b}$\\
PS1-i-PSFmag        & magnitude & {\it i}-band PSF magnitude$^{b}$\\
PS1-i-PSFmag-error  & magnitude & {\it i}-band PSF magnitude error$^{b}$ 
\enddata
\tablenotetext{a}{From TGSS-NVSS catalog \citep{2018MNRAS.474.5008D}}
\tablenotetext{b}{From PS1 photometric data \url
{http://mastweb.stsci.edu/ps1casjobs/home.aspx}}
\end{deluxetable*}

\subsection{Catalog description}

A description of the BROS catalog is summarized in Table \ref{tab:bros_ps1}. The catalog is available at \url{http://www.bao.city.ibara.okayama.jp/BROS/} and the Centre de Données astronomiques de Strasbourg \url{https://cds.u-strasbg.fr/}.

\begin{figure*}[!htb]
  \centering
  \includegraphics[angle=0,width=15cm]{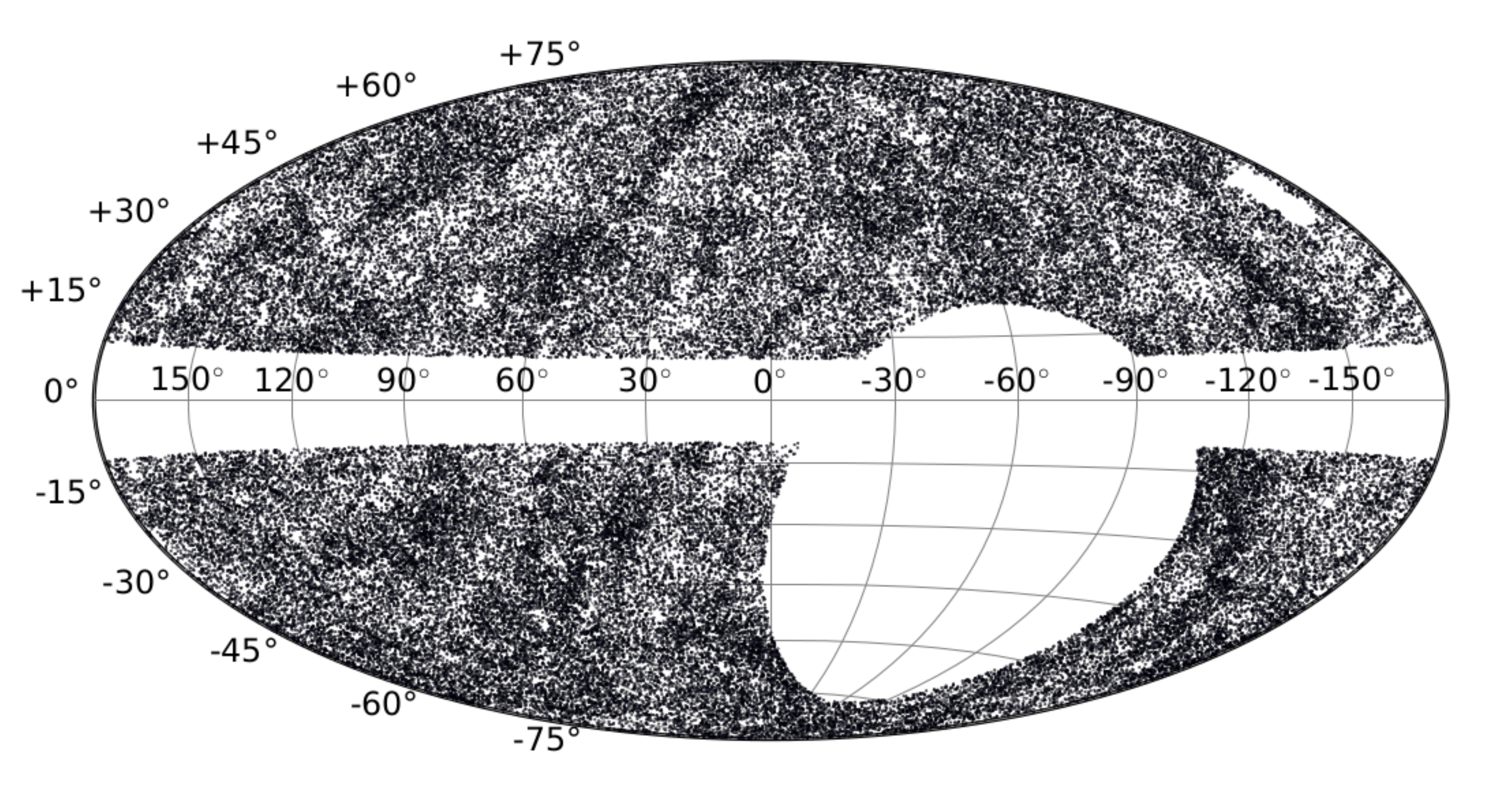}
  \caption{Location of all the 88,211 BROS sources in Galactic coordinates.
    The large elliptical region centered at $l \sim -60^{\circ}$ and $b \sim -30^{\circ}$
    corresponds to the southern sky of $\delta < -40^{\circ}$,
    where the NVSS does not cover.
    The small blank region located around $l \sim -150^{\circ}$
    and $b \sim 40^{\circ}$ are due to lack of TGSS sources in
    that region.}
  \label{fig:galmap}
\end{figure*}

Figure \ref{fig:galmap} shows the sky distribution of BROS sources.
The distribution is not perfectly uniform. 
A higher source density region (strip) is present, for example,
at (l, b) = ($0^{\circ}, 90^{\circ}$) to (l, b) = ($60^{\circ}, 10^{\circ}$).
We think that this is not intrinsic, but instead due to the nonuniform sensitivity
of the TGSS (i.e., the nonuniform distribution of TGSS sources),
as evidenced by the spatial distribution of TGSS sources in Fig. B.2 of \cite{2017A&A...598A..78I}.

\subsection{Comparison with other blazar catalogs}

CRATES is a well-known blazar candidate catalog \citep{2007ApJS..171...61H}; here, they selected radio sources of $S_{4.8 {\rm GHz}} \geq 65$ mJy showing flat radio spectra of $\alpha \geq -0.5$ between 1.4~GHz (or 0.84~GHz for the sources located at $\delta < -40^{\circ}$) and 4.8 GHz.  CRATES has 11,131 sources and is, thus, a quarter size of BROS. One major difference with respect to  BROS is the radio frequencies. Given that CRATES utilizes the NVSS catalog  for 1.4~GHz and GB6, PMN, and S5 catalogs for 4.8 GHz, the detection limit of CRATES depends on its declination with a typical detection limit at $\delta < +75^{\circ} \sim 65$ mJy at 4.8~GHz. Thus, the CRATES catalog is shallower than BROS.

The Candidate Gamma-Ray Blazar Survey catalog \citep[CGRaBS][]{2008ApJS..175...97H} is essentially based on the CRATES and contains 1,625 sources with radio and X-ray data. CGRaBS also contains red-shift and blazar-type classifications (FSRQ or BL Lac) based on optical spectroscopy. The Roma-BZCAT catalog \citep{2015Ap&SS.357...75M} contains 3,561 blazars selected by multi-frequency surveys. In the Roma-BZCAT, most of the blazars are classified into two types: flat-spectrum radio quasars (FSRQs) and BL Lac objects (BL Lacs), by optical spectroscopic data. 3HSP catalog \citep{2019A&A...632A..77C} contains 2,013 high synchrotron peaked blazar candidates made for high energy astrophysics and neutrino astronomy. CGRaBS, Roma-BZCAT and 3HSP focus on multi-wavelength observations and are useful for identifying blazar properties; however, the number of sources is smaller than that of BROS. 

Table \ref{tab:catalogs} shows a summary of blazar catalogs. Hereinafter, we focused on a comparison with CRATES, which adopts similar selection criteria to BROS. We cross-matched the BROS and CRATES catalogs. First, we selected CRATES sources located at $\delta > -40^{\circ}$; the resulting CRATES source number was 9,071. For these sources, we searched for the TGSS counterpart. Then, we identified 7,971 sources (88\%) with TGSS-NVSS counterparts; additionally, 5,178 sources (57\%) have BROS counterparts. The sources that have TGSS-NVSS counterparts but are not listed in the BROS catalog, are due to the soft radio spectra of $\alpha < -0.6$ or have a non source classification flag of 'S' or a radio compactness of $C > 1.6$.

\begin{figure}[!htb]
  \centering
  \includegraphics[angle=0,width=8cm]{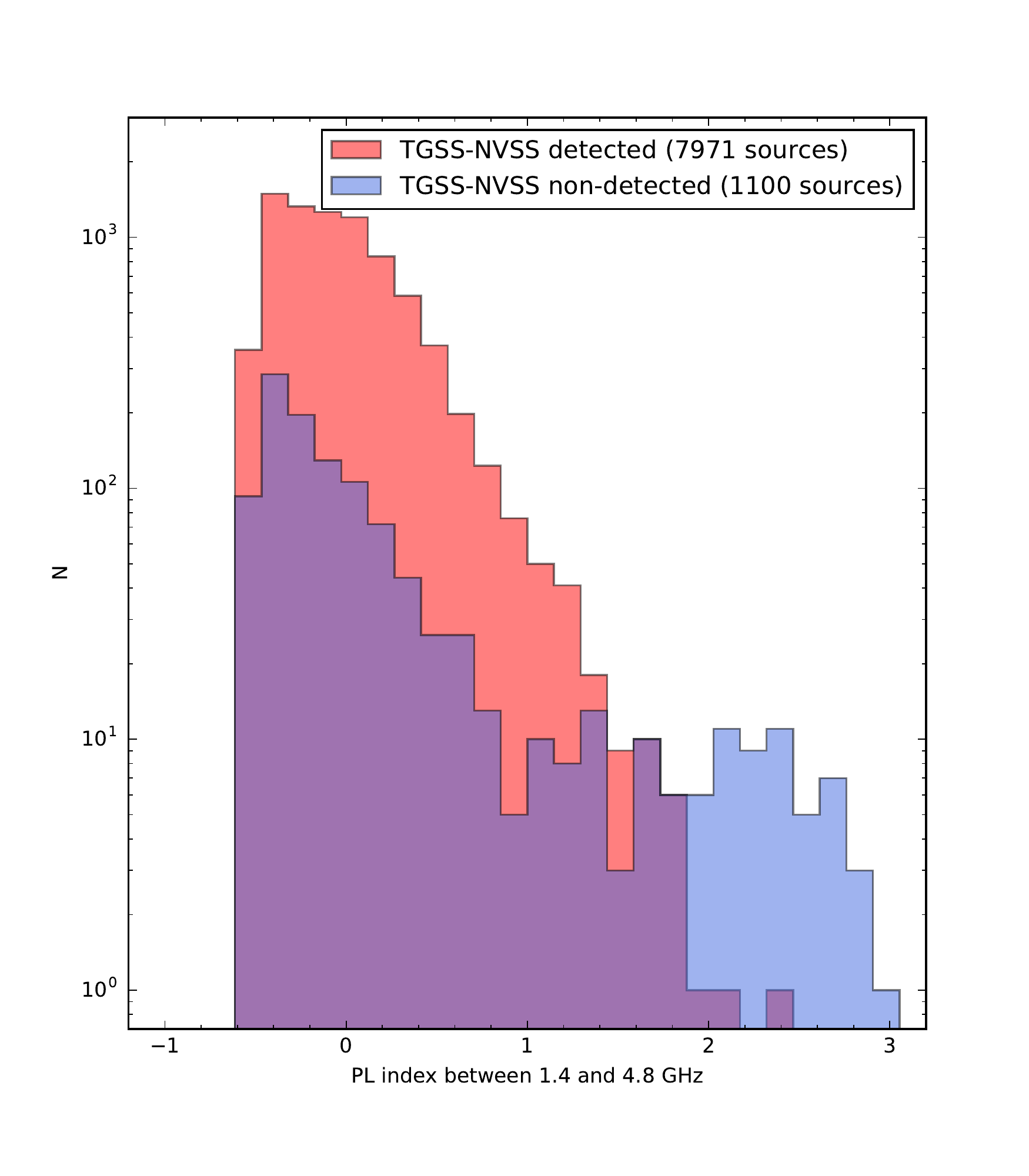}
  \caption{Distribution of power-law index between 1.4 and 4.8 GHz
    for CRATES sources. Shown in red (blue) are CRATES sources
    detected (not detected) in the TGSS-NVSS catalog. 
    }
  \label{fig:crates_alpha}
\end{figure}

We also checked the reason why about 12\% of the CRATES sources are not detected by the TGSS-NVSS catalog. One reason would be due to the sensitivity of the TGSS-NVSS catalog at 0.15~GHz. The typical RMS noise is 8 mJy beam$^{-1}$ at 0.15~GHz \citep{2018MNRAS.474.5008D}; sensitivity is not uniform. In the CRATES catalog, 4.8~GHz fluxes of the sources are larger than 65 mJy. Thus, if a source has a very hard spectrum of say 0.6, the source cannot be detected by the TGSS. To check this possibility, we made a histogram of $\alpha_{1.4-4.8 {\rm GHz}}$ for CRATES sources that are listed or dropped in the TGSS catalog (Figure \ref{fig:crates_alpha}). We find that the CRATES sources not detected by the TGSS-NVSS show a relatively hard spectral index compared to the TGSS-CRATES sources.
In addition, contamination of GPS/CSS sources in the CRATES catalog may also be a reason for non-detection sources at 0.15~GHz, because it has a spectral break peaking for the GHz band.

\begin{figure}[!htb]
  \centering
  \includegraphics[angle=0,width=8cm]{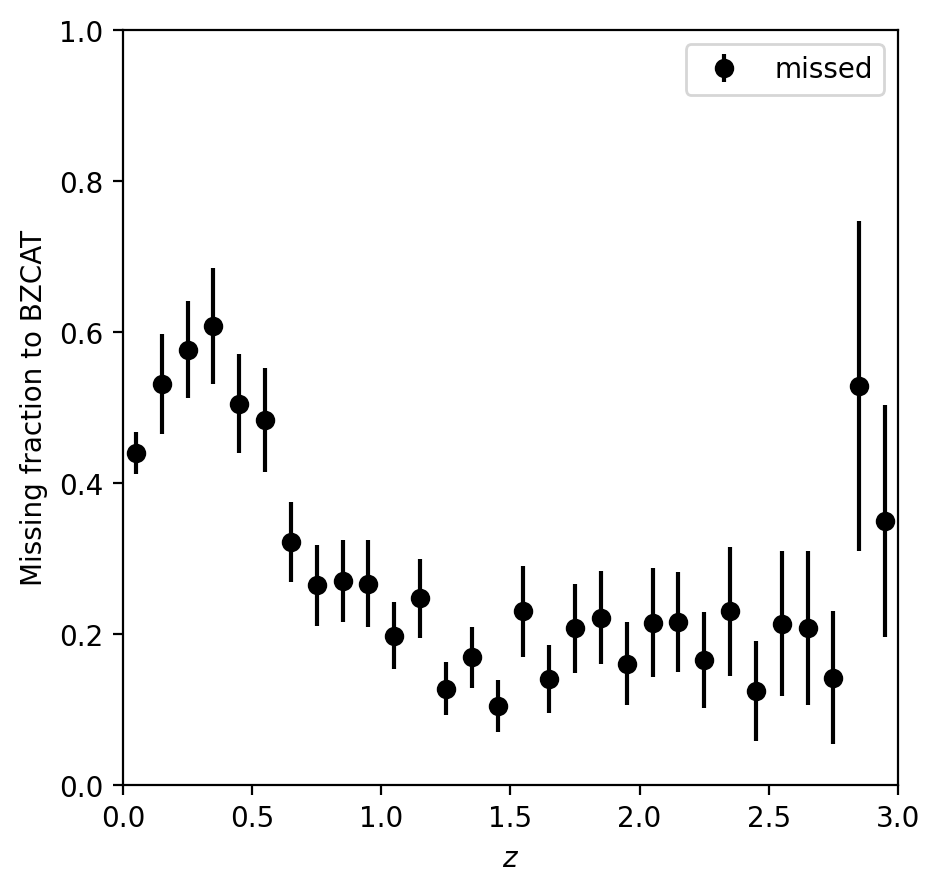}
  \caption{Missing ratio of BROS sources to Roma-BZCAT as a function of redshift}
  \label{fig:Completeness_BZCAT}
\end{figure}

1861 Roma-BZCAT sources have BROS counterparts within the BROS sky region, in other words, 1253 ($\sim40$\%) of the Roma-BZCAT objects are missed in the our catalog. Among the missing Roma-BZCAT sources, 67\% are recorded as either of 'BL Lac', 'BL Lac Candidate', 'BL Lac-galaxy dominated', 'Blazar Uncertain type', while the remaining objects are classified as 'QSO RLoud flat radio sp'. Figure \ref{fig:Completeness_BZCAT} shows the redshift distribution of the missing ratio of the Roma-BZCAT objects in BROS. The bump at low redshift ($z\lesssim0.5$) in the Figure \ref{fig:Completeness_BZCAT} is mostly dominated by the BL Lac sources, while at $z>0.5$ the missing population is dominated by FSRQs. The redshift measurements of BL Lac sources are known to be biased towards low redshifts $\sim0.5$ because of spectroscopic limitations \citep[see ][Figure 7 for detail]{2014ApJ...780...73A}. We further investigate the reason for the differences between BROS and Roma-BZCAT. About 70\% sources of the missed Roma-BZCAT sources are not listed  or having source classification flag of "L", i.e., non-TGSS detection, \citep[see ][for detail]{2018MNRAS.474.5008D} in the TGSS-NVSS catalog. By estimating the 0.15 GHz flux, we confirm that the corresponding flux of the missed Roma-BZCAT sources are below the detection limit of TGSS, where we extrapolate from the NVSS flux assuming a spectral index of 0.5. Therefore, we conclude the TGSS flux limit causes this mismatch. The remaining 30\% are excluded by its radio morphology.

The similar situations are also realized when comparing the BROS catalog with the 3HSP catalog  (363 missing sources, 21.5\%) and 4LAC (784 missing sources, 32\%). The BROS catalog is uniform over the footprint, but tends to miss the sources that have hard radio spectrum limited by TGSS sensitivity at 0.15~GHz.

\begin{deluxetable*}{lrcll}[t]
 \tablecaption{Comparison with other catalogs \label{tab:catalogs}}
 \tablecolumns{5}
 \tablehead{
 \colhead{Catalog$^{a}$} & \colhead{Sources$^{b}$} & \colhead{Sky coverage$^{c}$} & \colhead{band$^{d}$} & \colhead{Classification$^{e}$} 
 }
\startdata
BROS       & 88,211 & $\delta>-40^{\circ}$   & 0.15~GHz, 1.4 GHz, optical g, r, i-band & no \\
CRATES     & 11,131 & all sky  & 1.4 GHz, 8.4 GHz &  no\\
CGRaBS     & 1,625  & all sky & 8.4 GHz, optical R-band & yes\\
Roma-BZCAT & 3,561  & all sky & 1.4 GHz, 143 GHz, optical R-band, X-ray, gamma-ray& yes \\
3HSP & 2,013 & all sky & 1.4 GHz and X-ray & yes \\
\enddata
\tablenotetext{a}{Catalog name}
\tablenotetext{b}{Number of sources}
\tablenotetext{c}{Sky coverage. Note that all catalogs exclude $|b|<10^{\circ}$ region}
\tablenotetext{d}{Available band in the catalog}
\tablenotetext{e}{Blazar type classification}
\end{deluxetable*}

\section{Discussion}
\subsection{Optical properties}
\label{sec:dis_opt}

\begin{figure}[!htb]
  \centering
  \includegraphics[angle=0,width=8cm]{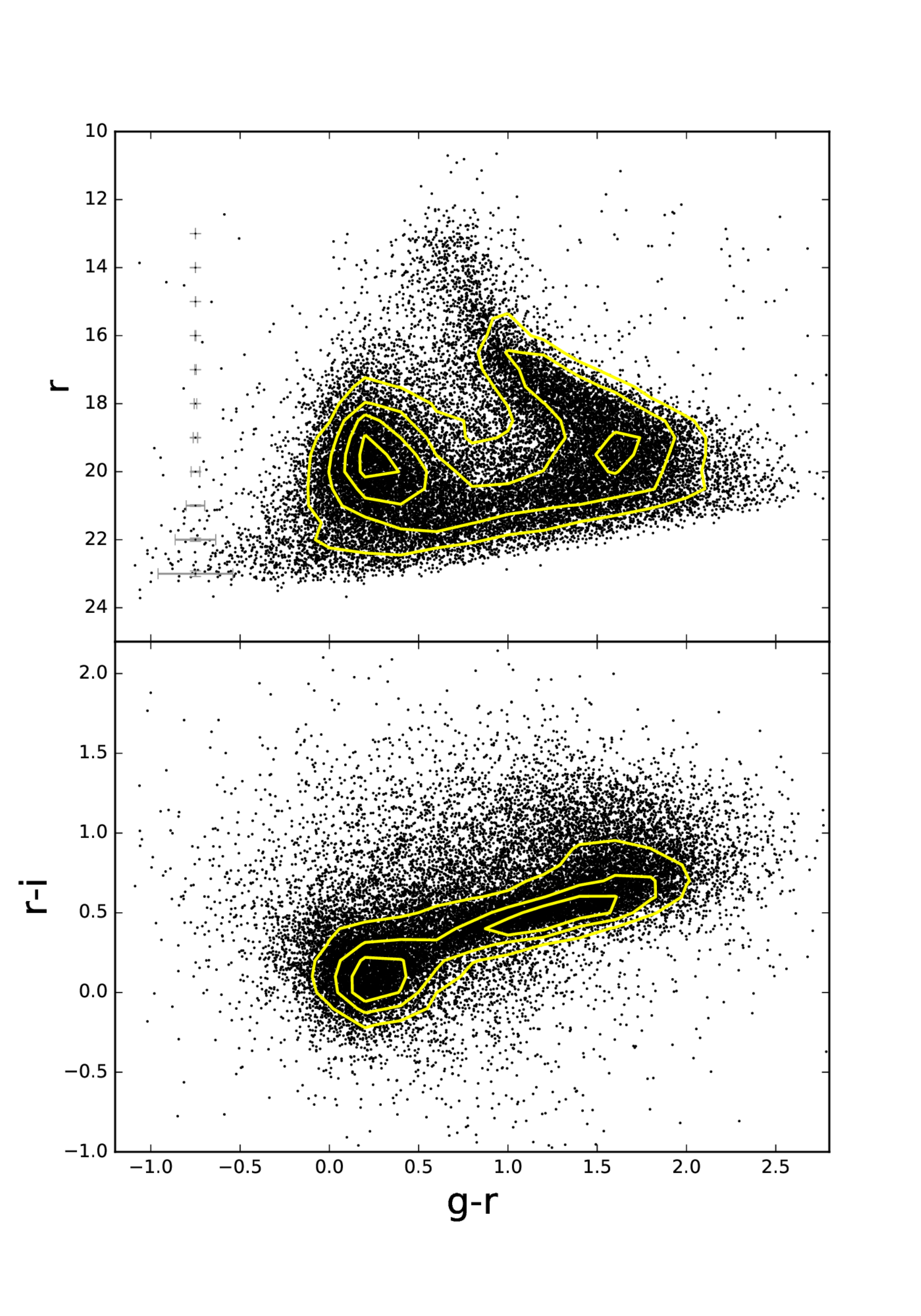}
  \caption{$g-r$ vs $g$ (upper) and $g-r$ vs $r-i$ (lower) plots for the BROS sources whose optical counterparts are detected by the PS1 survey. Cross gray data point indicates typical uncertainties of $g$-band magnitude and $g-r$ value.}
  \label{fig:bros_ps1_HR}
\end{figure}

Here we discuss the optical properties of BROS sources. Figure \ref{fig:bros_ps1_HR} upper and lower panel shows the $g-r$ vs. $r$ color-magnitude and $g-r$ vs. $r-i$ color-color diagrams for the BROS sources, respectively. Magnitudes are in Kron magnitude, which is photon flux measured with the Kron technique for an extended source \citep{1980ApJS...43..305K}, because many BROS sources show extended optical emission, i.e., host galaxies. We exclude sources with large photometric uncertainties ({\it g}-error, {\it r}-error or {\it i}-error $>$ 0.3 mag.) from the plot.

The upper panel in Fig. \ref{fig:bros_ps1_HR} clearly shows the presence of two populations: one cluster at  $(g-r, r) \sim (0.2, 18.0)$ to $(0.2, 22.0)$ and the other at  $(g-r, r) \sim (0.8, 18.0)$ to $(1.6, 22.0)$. The latter population is not seen in CRATES catalog (see Appendix A for details about the CRATES-PS1 sources and their comparison with BROS-PS1 sources).  The horizontally elongated distribution at $(g-r, r) \sim (0-1.8, 22.0)$ shown in the upper panel is not intrinsic, but is due, instead, to the large error associated with the detection limit of the PS1 catalog. 

\begin{figure}[!htb]
  \centering
  \includegraphics[angle=0,width=8cm]{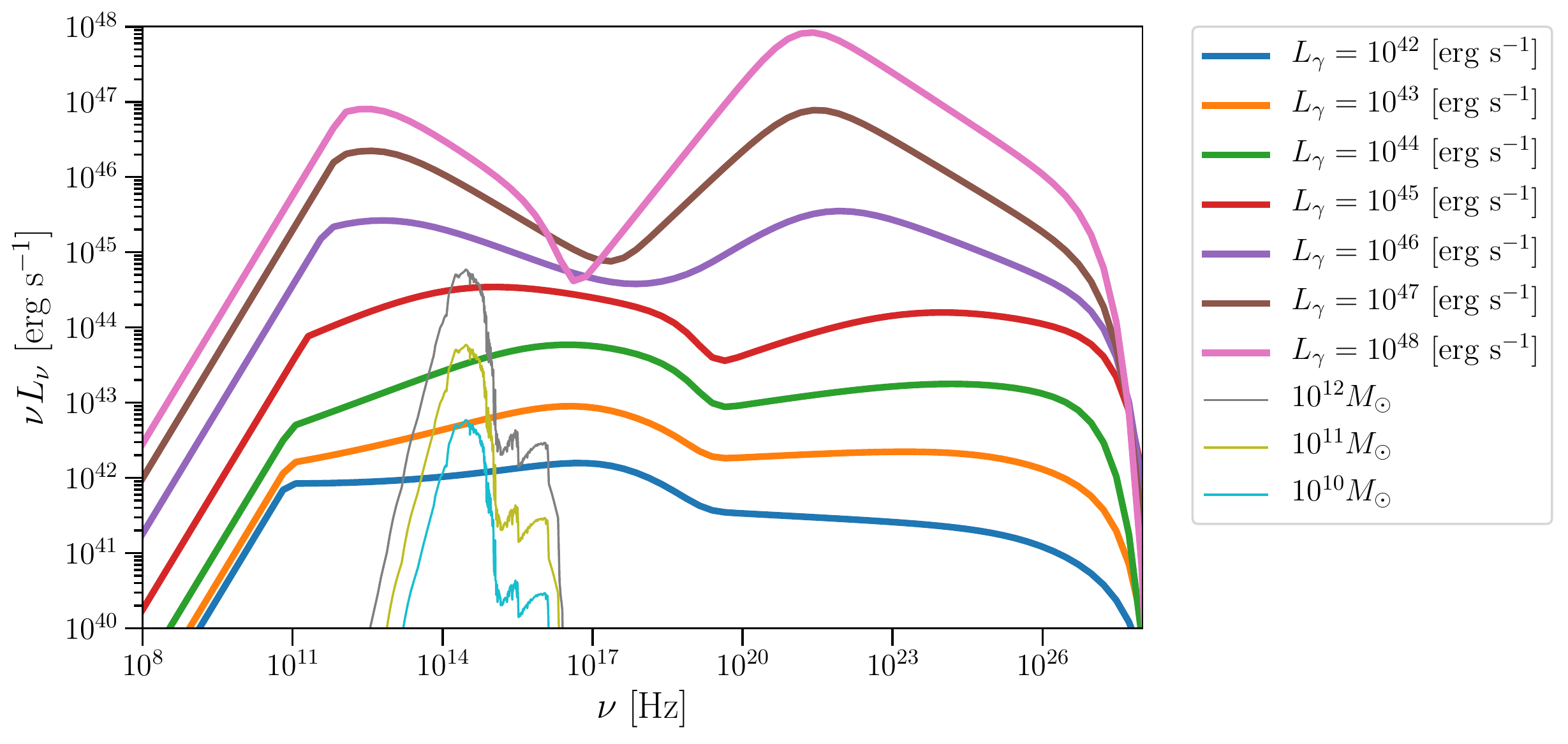}
  \caption{Overall SED of blazar with host galaxy templates with various stellar mass and jet luminosity as indicated in the figure.}
  \label{fig:SED_blazar}
\end{figure}

To understand the nature of these two populations, we calculated the expected {\it gri} magnitudes and colors, utilizing phenomenological spectral energy distribution (SED) templates of a composite of blazars and elliptical galaxies. Blazar's broadband SED consists of two spectral peaks (synchrotron and, most likely, inverse Compton emissions) and that the fainter blazars show higher peak frequency due to inefficient radiative cooling, so-called blazar sequence \citep[e.g.,][]{1998MNRAS.299..433F,1998ApJ...504..693K,Ghisellini2017}. We adopted the latest blazar sequence model in \citet{Ghisellini2017}. As described later, this model well explained its magnitude-color diagrams for the BROS sources. However, we note that the existence of the blazar sequence is still under debate because of selection effect \citep[e.g.,][]{2004MNRAS.348..937C,2007ApJ...662..182P,2012MNRAS.420.2899G,2015MNRAS.450.2404G,2019MNRAS.484L.104P}. For elliptical galaxies, we used the stellar population synthesis model by \citet{Bruzual2003} assuming the Salpeter initial mass function, instantaneous star formation, age of 10~Gyr old, and solar metallicity. Figure \ref{fig:SED_blazar} shows the overall SED with bright host galaxy SED templates of various stellar mass.

\begin{figure}[!htb]
  \centering
  \includegraphics[angle=0,width=10cm]{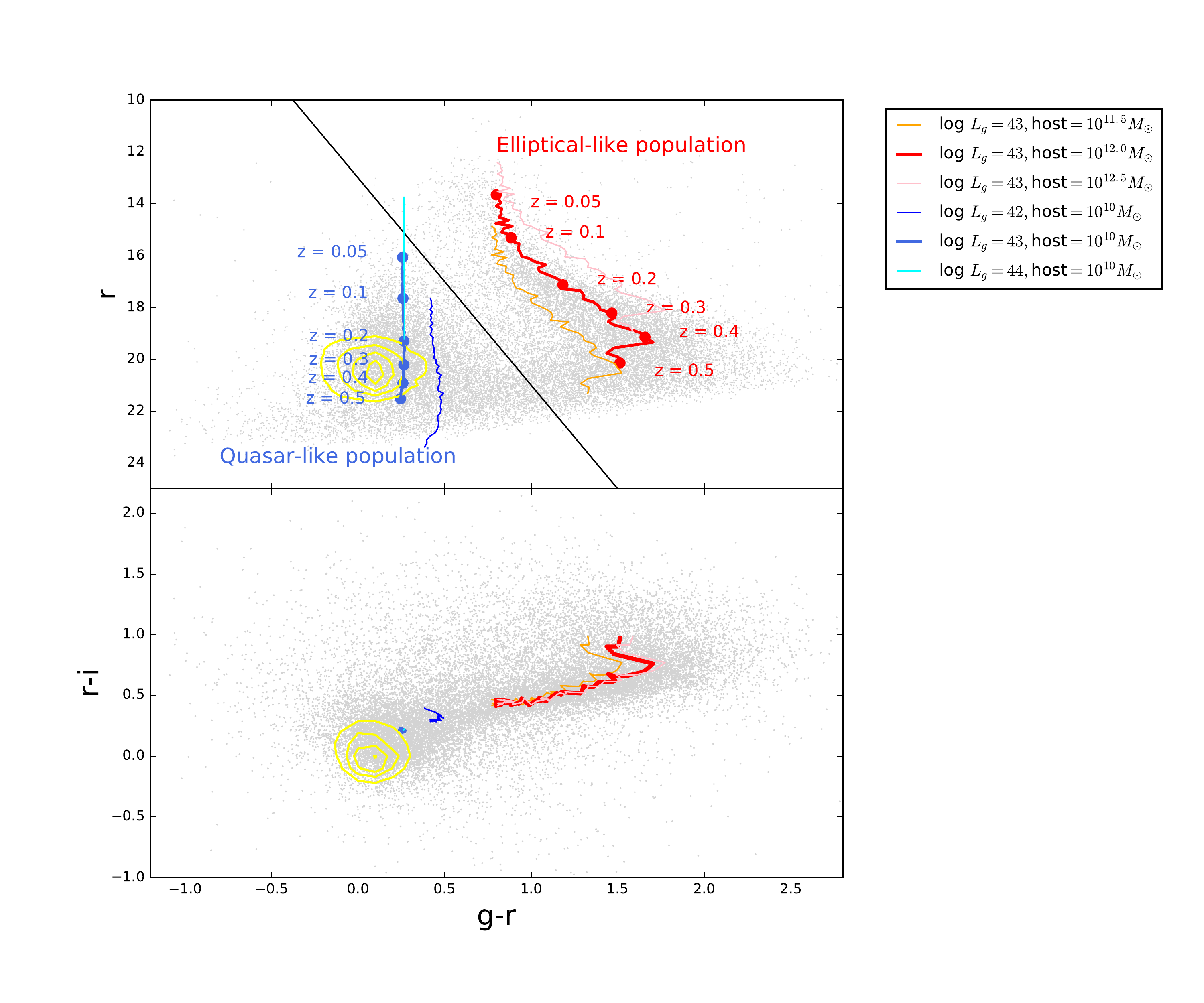}
  \caption{Same as Figure \ref{fig:bros_ps1_HR}, but color and magnitude calculated using the elliptical galaxy template with  around $10^{12} M_{\odot}$ and blazar SED template of around $L = 10^{43}$ at z = 0.05 to z = 0.5 are overlaid in red. Also, calculated using elliptical galaxy template with around $10^{10} M_{\odot}$ and blazar SED templates with luminosity of around $10^{44}$ are overlaid in blue. Yellow contour indicates the normal quasars located at $z< 1.0$ \citep[SDSS quasars,][]{2017A&A...597A..79P}. Details are described in text.}
  \label{fig:bros_ps1_HR_temp}
\end{figure}

Figure \ref{fig:bros_ps1_HR_temp} shows the $g-r$ vs $r$ color-magnitude and $g-r$ vs $r-i$ color-color diagrams of blazars with several gamma-ray luminosity $(L_{g}$ and host galaxy stellar mass $M_{\rm host})$, at various redshifts.  BROS sources are also shown. The model curves based on our simple calculation closely reproduce the observed population, particularly $(L_{g}, M_{\rm host})=(10^{43}~{\rm erg\ s^{-1}}, 10^{12}M_\odot)$ and $(10^{44}~{\rm erg\ s^{-1}}, 10^{10}M_\odot)$. The optical emission is dominated by bright hosts in the former model, whereas in the latter case, the emission is dominated by blazars. Thus, the right population is dominated by bright elliptical galaxy ($z < 0.5$) emission, whereas the left population is dominated by blazar emission.
Large samples (13,153 sources, 52\%) which have bright host galaxy are newly discovered by BROS, while hundreds of similar samples are reported in Roma-BZCAT, 3HSP and other catalogs (e.g., 1,207 sources, 19\% are classified as latter population in CRATES catalog, see Appendix A). We also note that 'big blue bump' component originated in the disk is confirmed in several FSRQs in the optical band. For reference, the colors-magnitude and color-color diagram of normal SDSS quasar located at low redshift ($z < 1$) are shown in Figure \ref{fig:bros_ps1_HR_temp} as yellow contour \citep{2017A&A...597A..79P}. As we see in the figure, the observed low redshift SDSS quasars which do not deviate from the simple Blazar model confirm that the contribution from the disk component is indistinguishable in our distinction. 

Hereinafter, for the sake of simplicity, we divided the BROS sources into two populations; namely quasar-like ($r > 8 \times (g-r) + 13$  in $g-r$ vs $r$ plot) and elliptical-like populations ($r < 8 \times (g-r) + 13$  in $g-r$ vs $r$ plot). In the BROS catalog, if the sources have a PS1 candidate and are not classified as GPS sources, we set 'Q' ('E') flag for the source which locates in a quasar-like (elliptical-like) population to the 'SourceFlag' column. The sources that do not have a PS1-candidate and are not classified as GPS sources were assigned a 'B' flag.

We note that there is also a marginal population in the valley between quasar-like and elliptical-like populations. This marginal population is likely a mixture of the distribution tails of both population.. Further detailed studies such as the convolution of blazar and galaxy luminosity functions will be able to elucidate the properties of the marginal population; however, this is beyond the scope of this paper.

We also investigated the spatial extension of sources in the optical band. We use the $iPSFMag - iKronMag > 0.05$ formula, where $iPSFMag$ is the photon flux derived with a point source spread function and $iKronMag$ is the photon flux for the extended source, to select the extended source. 
As a result, $\sim 93$\% elliptical-like sources were classified as extended sources; in contrast, $\sim 22\%$ quasar-like sources were classified as extended sources. This result also supports an "elliptical-like" branch, determined by its optical color, consisting of elliptical galaxies.

Generally, optical emission in FSRQs is not dominated by a host galaxy, as FSRQs have strong jet emission ($L_{g}> 10^{46}$ erg s$^{-1}$) compared to that of BL~Lac-type objects (see Fig. \ref{fig:SED_blazar}).
Therefore, it is expected that the elliptical-like population consists mostly of BL~Lacs, whereas the quasar-like population consists of both FSRQs and BL Lacs.  We also confirmed this by cross-matching between BROS and classified sources (BL Lac or FSRQ) in 4LAC. As a result, 74\% (26\%) of elliptical-like sources are classified as BL Lacs (FSRQs) in 4LAC, and 44\% (56\%) of quasar-like sources are classified as BL Lacs (FSRQs).

\subsection{Comparison to radio properties}

\begin{figure}[!htb]
  \centering
  \includegraphics[angle=0,width=8cm]{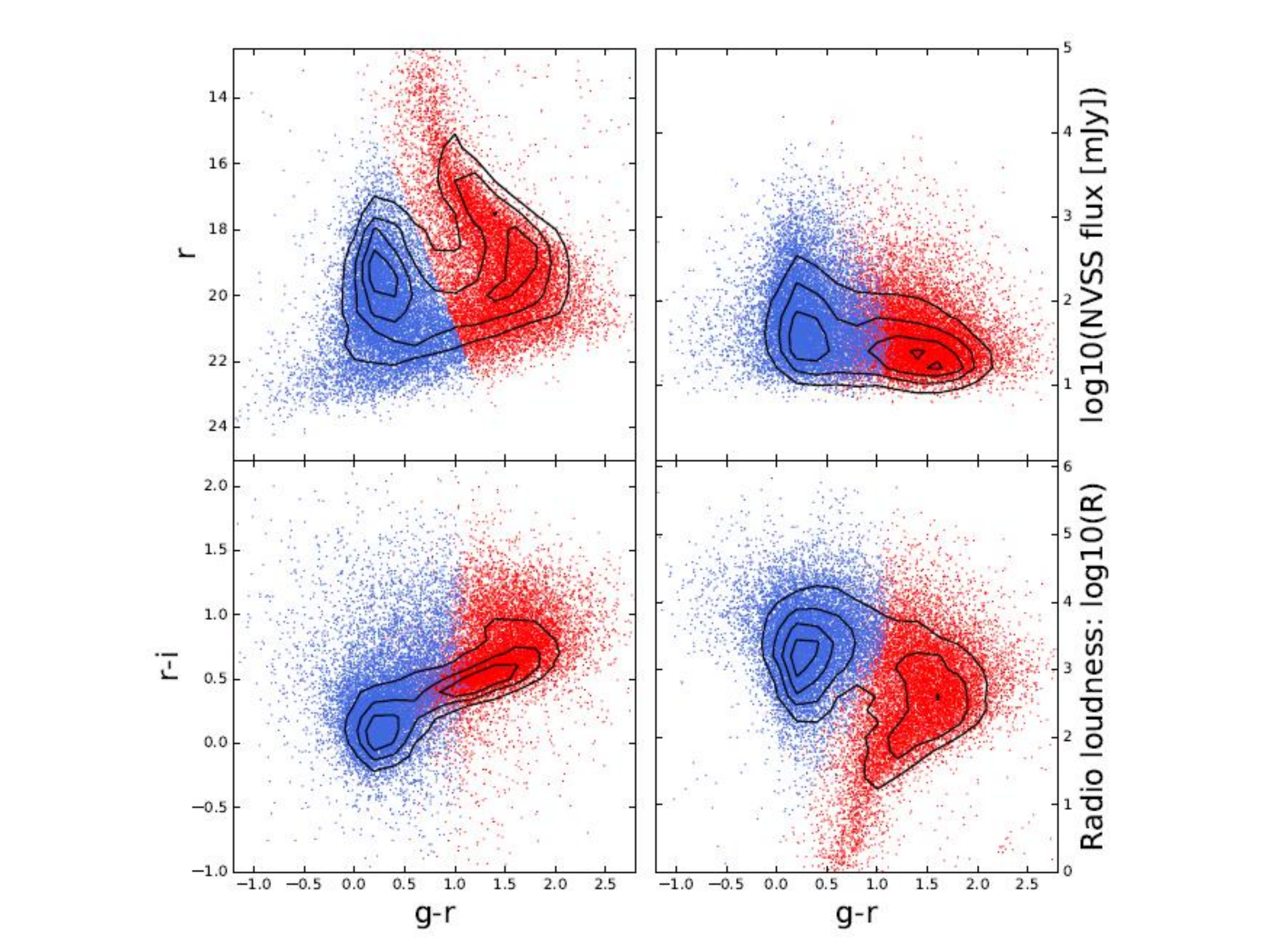}
  \caption{Color, radio flux, and radio-loudness for objects inside
    the quasar-like and elliptical-like regions,
    which are shown in blue and red, respectively.
    See Figure \ref{fig:bros_ps1_HR_temp} for the definition
    of each region.}
  \label{fig:bros_ps1_HR_sep_4}
\end{figure}

Figure \ref{fig:bros_ps1_HR_sep_4} shows the $g-r$ vs. NVSS flux and $g- r$ vs. radio loudness $R$ (defined as $F_{{\rm 1.4~GHz}}/F_{r}$) of the BROS sources together with the magnitude and color diagram. As shown in the upper right panel of Figure \ref{fig:bros_ps1_HR_sep_4}, quasar-like sources show slightly larger NVSS fluxes compared to the elliptical-like one. 

The quasar-like sources tended to have a high $R$ compared to that of elliptical-like ones. Radio loudness $R$ is a useful tool to distinguish the spectral type of blazar, as a high $R$ value is expected for BL~Lac-type objects from the unification model for AGN \citep[e.g.,][]{1995PASP..107..803U}.
However, in our sample, optical emission of elliptical-like sources are likely contaminated by the host galaxy, such that our $R$ value tends to be small. Therefore, we believe that the differences in $R$ between quasar-like sources and elliptical-like ones are not due to the spectral difference originating from jet emission.

\subsection{Cumulative source count distribution}
\begin{figure}[!htb]
  \centering
  \includegraphics[angle=0,width=8cm]{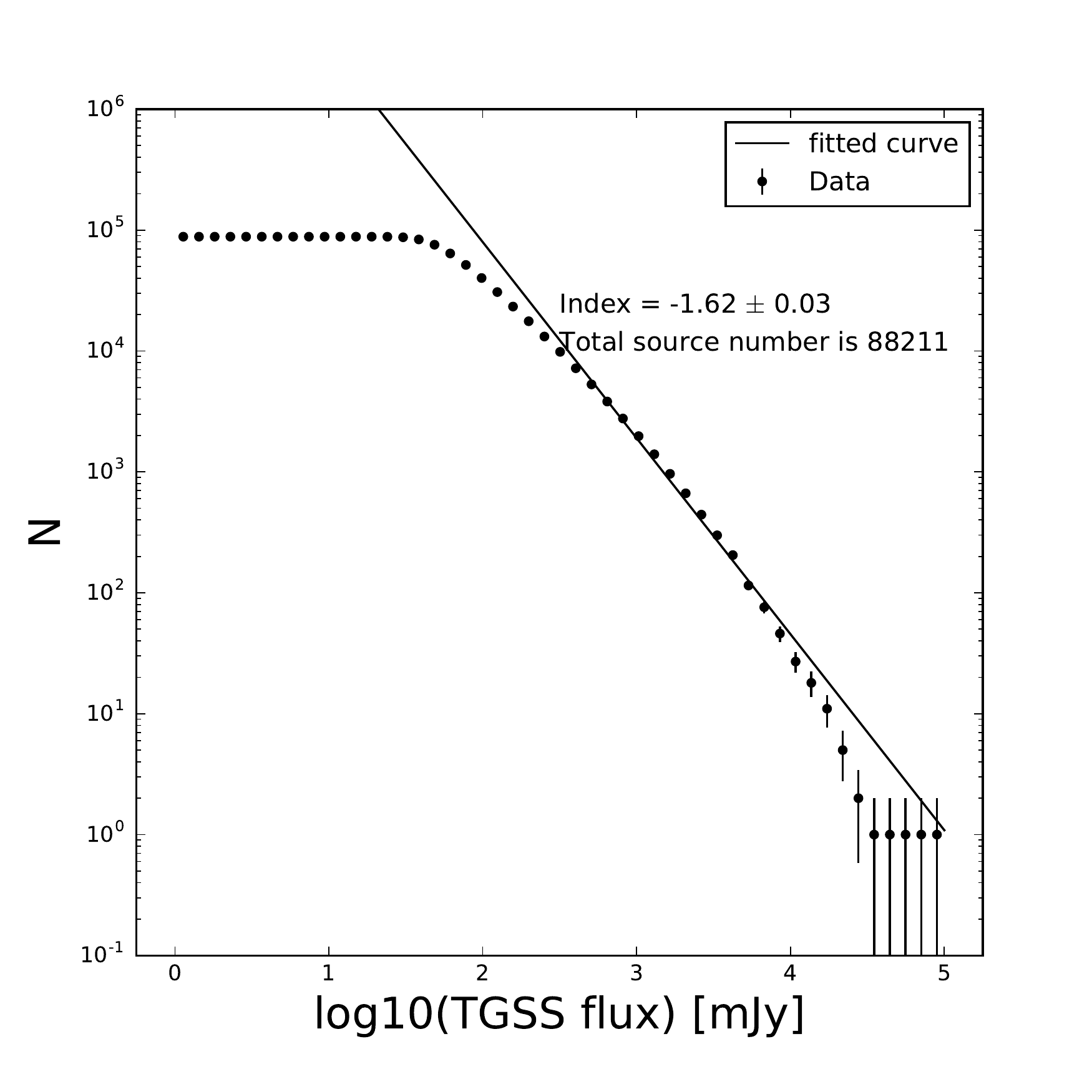}
  \includegraphics[angle=0,width=8cm]{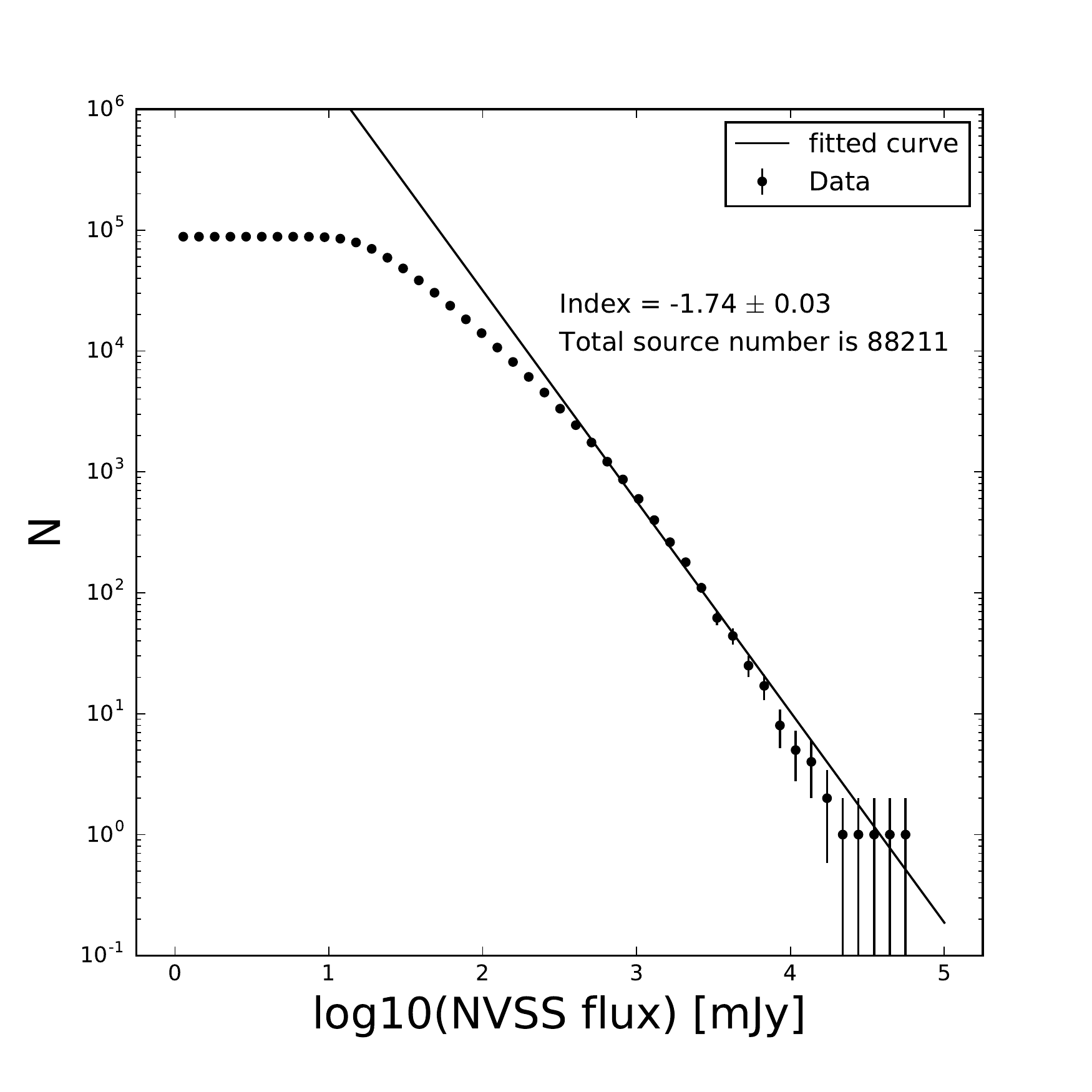}
  \caption{logN-logS plots for the BROS sources using the TGSS
    (left) and NVSS (right) fluxes.
    In both of the plots, fitting was performed for
    the data points above 500 mJy}
  \label{fig:logNlogS_all}
\end{figure}

Figure \ref{fig:logNlogS_all} shows the $\log N$-$\log S$ distributions for the BROS sources based on TGSS and NVSS fluxes. We fit the $\log N$-$\log S$ data points  using a power-law function of $N \propto S^{\beta}$. The NVSS-TGSS catalog is estimated to be complete down to $\sim 5$ mJy for NVSS and $\sim 100$ mJy for TGSS sources \citep{2018MNRAS.474.5008D}. To avoid the effects of source incompleteness for a lower radio flux, we used sources with $\ge500$~mJy for the derivation of the slope.
We obtain the power-law indices of $-1.62 \pm 0.03$ and $-1.74 \pm 0.03$ for TGSS and NVSS fluxes, respectively. The power-law index becomes $-1.5$ if the space density of objects is constant in Euclidean space \citep[e.g.,][]{1997iagn.book.....P}. The derived indices steeper than -1.5 indicate positive evolution for the selected flat-spectrum radio sources, namely the space density in the past is higher than that at present.

\begin{figure}[!htb]
  \centering
  \includegraphics[angle=0,width=8cm]{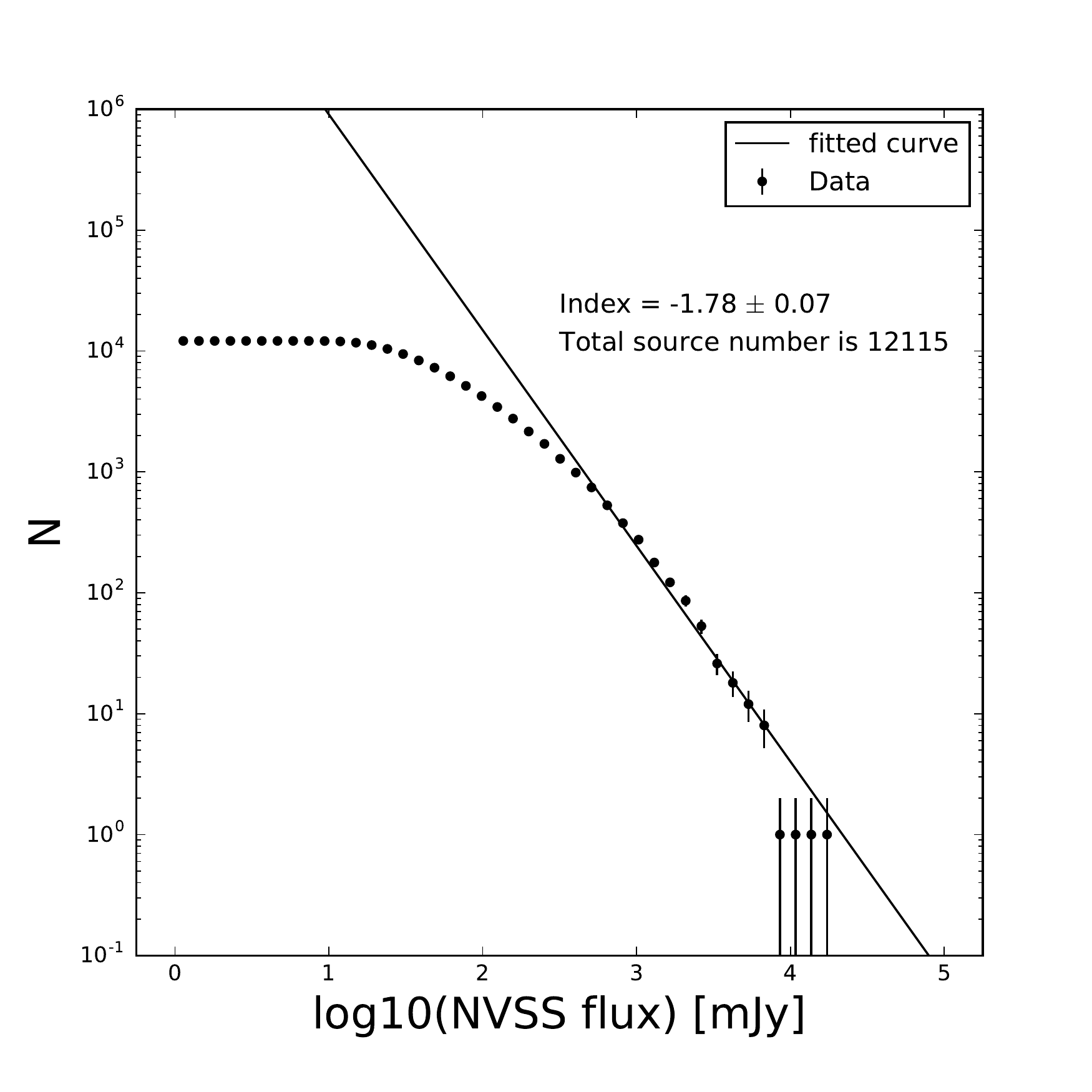}
  \includegraphics[angle=0,width=8cm]{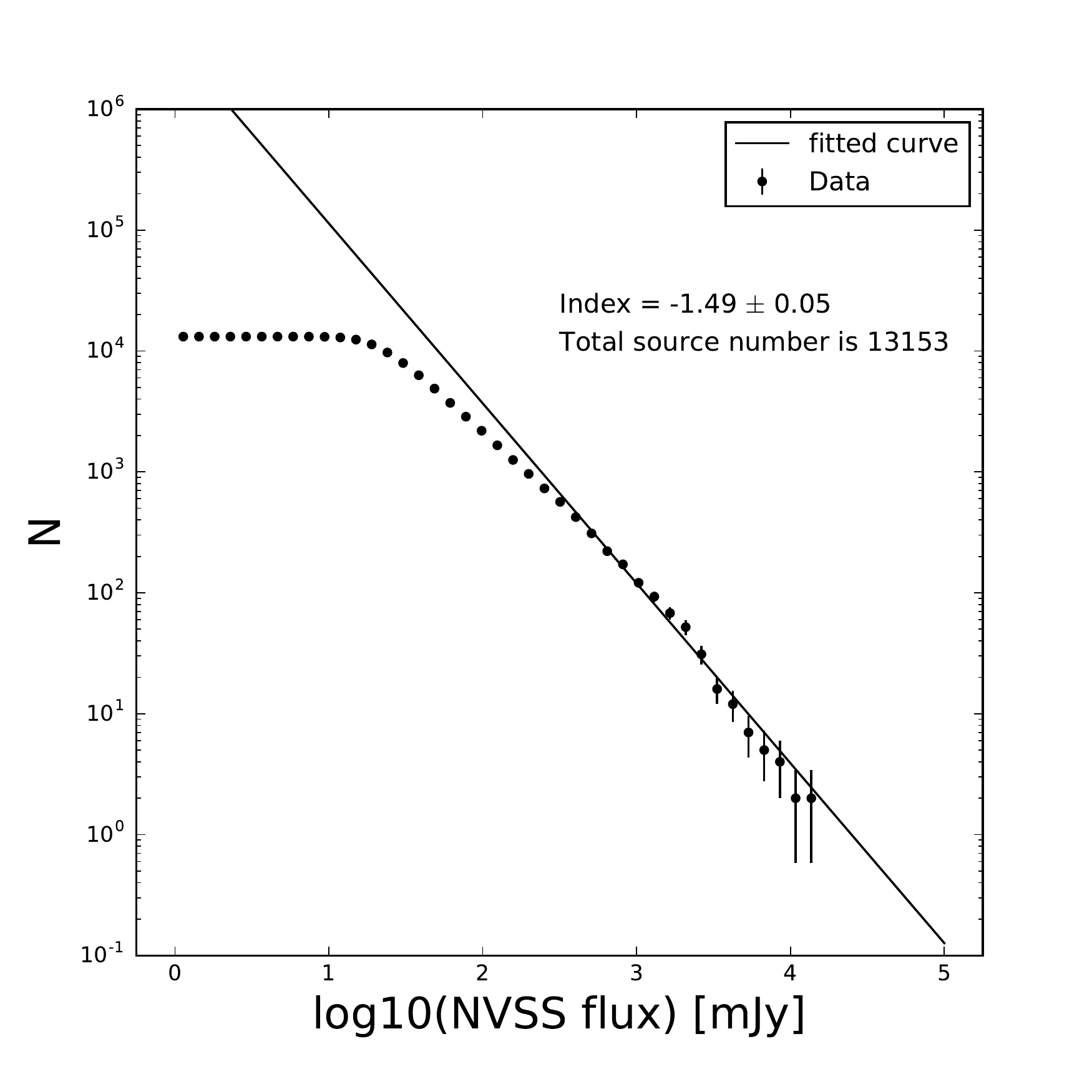}
  \caption{$\log N$-$\log S$ for quasar-like (left) and elliptical-like (right)
    BROS sources.
    Fit was performed for the data points above 500 mJy.}
  \label{fig:logNlogS_sep}
\end{figure}

The $\log N$-$\log S$ plots for the quasar-like and elliptical-like BROS sources are displayed in Figure \ref{fig:logNlogS_sep}. The quasar-like and elliptical-like sources show power-law indices of $-1.78 \pm 0.07$ and $-1.49 \pm 0.05$, respectively, indicating a significant difference between the two populations. The former population shows positive evolution whereas the latter does not, supporting our interpretation that the former is dominated by quasars and the latter consists of a different type of object relatively nearby elliptical galaxies.

\section{Conclusion}

Our main results are summarized as follows.
\begin{enumerate}
\item We produced a new blazar candidate catalog by selecting the flat-spectrum radio sources of $\alpha \ge -0.6$ from TGSS-NVSS catalogs; 88,211 sources are listed for the region of $\delta > -40^{\circ}$ and outside galactic plane ($|b| > 10^{\circ} $).
\item We identified the optical counterpart for each BROS source from Pan-STARRS1 photometric data; 42,757 sources are detected in the $r$-band.
\item The color-magnitude and color-color distribution of the BROS sources indicates the existence of two distinct populations in the BROS sources. One is a quasar-like population in which optical emission is dominated by blazars, and the other is an elliptical-like population in which optical emission is dominated by  bright elliptical galaxies located at $z \lesssim 0.5$. 
These populations roughly correspond to FSRQs and BL Lacs, respectively; however, jet-dominated BL Lac objects are included in the quasar-like population due to the blue synchrotron spectrum.
\item The BROS elliptical-like population includes nearby BL Lac objects, a fraction of which would be TeV emitters.
\end{enumerate}

Finally, we note that the BROS catalog is expected to be useful for identifying MeV/GeV unassociated sources detected by Fermi-LAT. For example, it is interesting to search for BROS sources inside the 4FGL unassociated sources. Further studies are needed to investigate the relationship between these two sources. In addition, very hard BL Lac objects, known as extreme HBLs, are considered faint in both optical and TeV gamma-ray bands, according to the blazar sequence. Indeed, for a famous extreme HBL 1ES~0229+200 and possible candidate HESS~J1943+213, the optical fluxes are dominated by host galaxy emission of bright ellipticals \citep{2009MNRAS.399L..59T,2014ApJ...787..155T}. Thus, we expect that numerous TeV sources are buried in the BROS elliptical-population, which makes them good targets for CTA observation.

\begin{acknowledgements}
We thank Masaomi Tanaka for fruitful discussions. This work was supported by the Grants-in-Aid for Scientific Research of the Japan Society for the Promotion of Science (18H03720, 16H01088). YU is supported by the U.S. Department of Energy under contract number DE-AC02-76-SF00515. YI is supported by JSPS KAKENHI Grant Number JP16K13813, JP18H05458, JP19K14772, program of Leading Initiative for Excellent Young Researchers, MEXT, Japan, and RIKEN iTHEMS Program. 
\end{acknowledgements}

\appendix
\section{CROSS MATCHING BETWEEN CRATES AND PS1 CATALOGS}

\begin{figure}[!htb]
  \centering
  \includegraphics[angle=0,width=7cm]{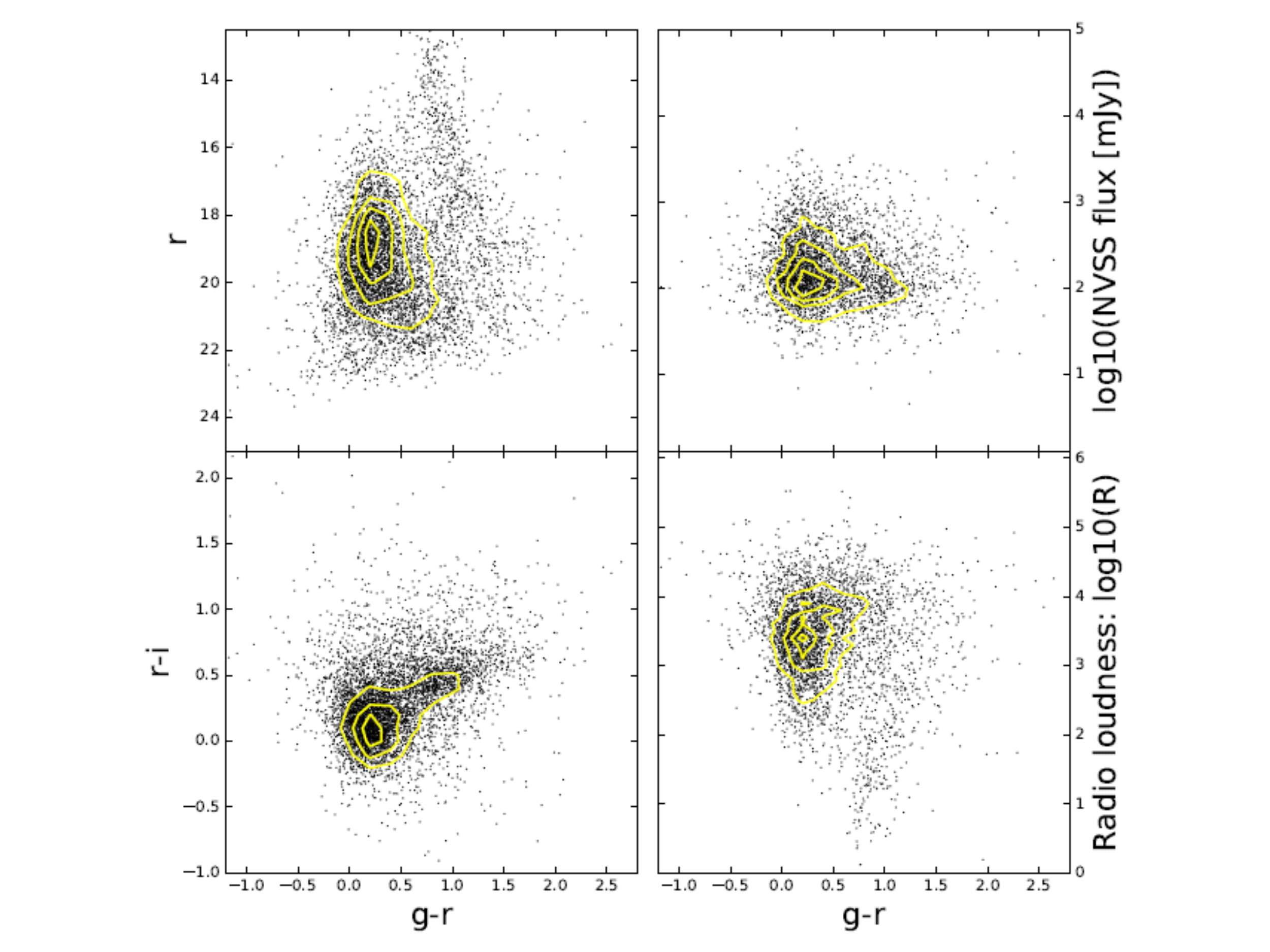}
  \caption{Same as Figure \ref{fig:bros_ps1_HR_sep_4} but for the
    CRATES sources whose optical
    counterparts are identified by the PS1 catalog.
    Note that the elliptical-like population detected in $g - r$ vs $r$
    plot for BROS sources (see Figure \ref{fig:bros_ps1_HR_temp})
    are barely seen and mostly undetected here.
    See upper left and lower left panels.}
  \label{fig:crates_ps1_HR}
\end{figure}

For comparison, we searched for the optical counterparts of CRATES
sources using PS1 photometric data.
The cross-matching method is completely the same as that utilized for BROS and PS1
cross-matching described in \S 3.
The results are shown in Figure \ref{fig:crates_ps1_HR}, which is the same as
Figure \ref{fig:bros_ps1_HR_sep_4} but for the CRATES-PS1 sources.
From the upper left and lower left panels, it is apparent that we mostly detected only objects in the BROS quasar-like population 
and few objects in the BROS elliptical-like population.

\begin{figure}[!htb]
  \centering
  \includegraphics[angle=0,width=7cm]{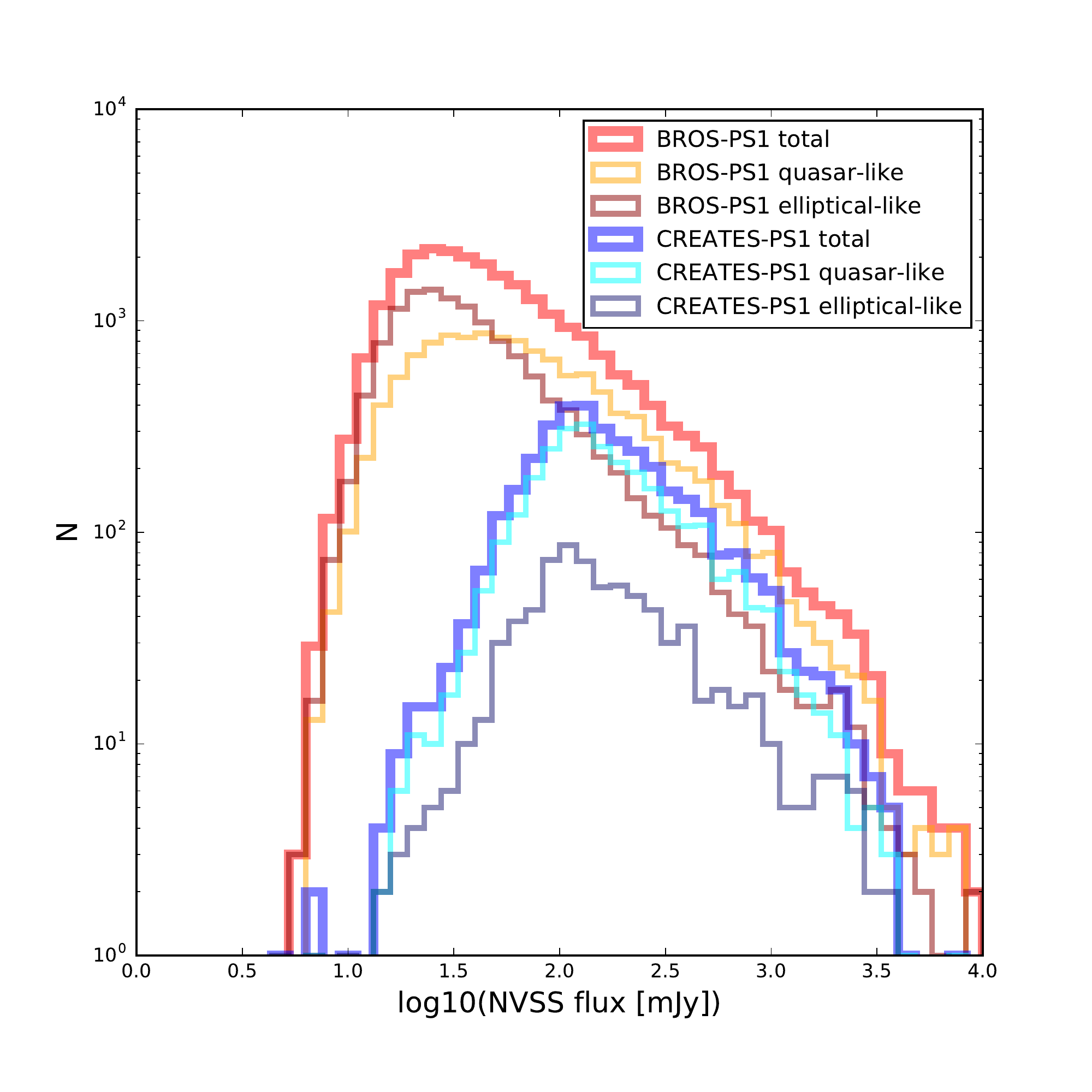}
  \caption{Histogram of NVSS fluxes for BROS-PS1 sources and CRATES-PS1
    sources in quasar-like (yellow: BROS-PS1, aqua: CRATES-PS1),
    elliptical-like (brown: BROS-PS1, darkblue: CRATES-PS1) population and 
    histogram for total number of BROS-PS1 (red) and CRATES-PS1 (blue).}
    \label{fig:hist_NVSS_sep}
\end{figure}

The reason why this CRATES-PS1 cross-matching does not pick up the
objects in the elliptical-like population would be that this population 
shows systematically lower radio fluxes than those in the
quasar-like population (see upper right panel in Figure \ref{fig:bros_ps1_HR_sep_4}).
To investigate this hypothesis in detail, we made histograms of NVSS
fluxes for CRATES-PS1 and BROS-PS1 sources in quasar-like and
elliptical-like populations, which are plotted in Figure \ref{fig:hist_NVSS_sep}.
The classification method for CRATES-PS1 sources into quasar-like
and elliptical-like populations is the same as that shown in Figure \ref{fig:bros_ps1_HR_sep_4}.
Clearly, the CRATES sources are the brightest portion of the BROS
sources in the NVSS flux and are mostly dominated by quasar-like
radio-bright objects.
Indeed, the radio flux threshold for the CRATES selection criteria
is relatively high (see also Figure \ref{fig:hist_NVSS_sep} for 
the NVSS flux threshold).
Thus, most objects in the BROS elliptical population are not
picked up in the CRATES catalog.

\bibliographystyle{apj}
\bibliography{bibbros}

\end{document}